\documentclass[english,reprint,amsmath,amssymb,aps,pra,superscriptaddress]{revtex4-1}
\usepackage[latin9]{inputenc}
\usepackage{amsmath}
\usepackage{graphicx}
\usepackage{esint}

\makeatletter
\usepackage{dcolumn}
\usepackage{bm}

\makeatother

\usepackage{babel}
\begin{document}

\title{Electron trimer states in conventional superconductors}

\author{Ali Sanayei}
\email{asanayei@physnet.uni-hamburg.de}

\address{Zentrum f{\"u}r Optische Quantentechnologien and Institut f{\"u}r
Laserphysik, Universit{\"a}t Hamburg, Luruper Chaussee 149, D-22761
Hamburg, Germany}

\author{Pascal Naidon}
\email{pascal@riken.jp}

\address{RIKEN Nishina Centre, RIKEN, Wak\={o} 351-0198, Japan}

\author{Ludwig Mathey}
\email{lmathey@physnet.uni-hamburg.de}

\address{Zentrum f{\"u}r Optische Quantentechnologien and Institut f{\"u}r
Laserphysik, Universit{\"a}t Hamburg, Luruper Chaussee 149, D-22761
Hamburg, Germany}

\address{The Hamburg Centre for Ultrafast Imaging, University of Hamburg,
Luruper Chaussee 149, D-22761 Hamburg, Germany}

\date{\today}
\begin{abstract}
We expand the Cooper problem by including a third electron in an otherwise
empty band. We demonstrate the formation of a trimer state of two
electrons above the Fermi sea and the third electron, for sufficiently
strong inter-band attractive interaction. We show that the critical
interaction strength is the lowest for small Fermi velocities, large
masses of the additional electron and large Debye energy. This trimer
state competes with the formation of the two-electron Cooper pair,
and can be created transiently via optical pumping.
\end{abstract}
\maketitle
In a seminal paper, Ref. \cite{Cooper_seminal_paper}, L.N. Cooper
showed that two electrons immersed in a Fermi sea form a bound state
for arbitrarily weak attractive interactions. The Cooper problem assumes
the dominance of the effective interaction induced by the electron-phonon
interaction over the screened Coulomb potential \cite{BCS_original_paper,Abrikosov,Schrieffer_book,Tinkham}.
It is modeled as constant in momentum space within a narrow energy
range of the order of the Debye energy for the relative kinetic energy
of the electrons. This simplified model distills the key features
and energy scales of the full interaction induced by electron-phonon
coupling that are relevant for the formation of the bound state and
its properties. The existence of this bound state indicates that the
noninteracting Fermi sea is unstable against pair formation, which
suggests the emergence of a superconducting state. A more extensive
theory of this state was provided by BCS theory \cite{BCS_original_paper,Abrikosov,Schrieffer_book,Tinkham,Aschroft,Baym,Fetter_Walecka,Lifshitz,Leggett,Grosso},
which elaborated on the essential ingredients that are necessary for
the formation of conventional superconductors, pointed out by the
Cooper problem and its solution.

\begin{figure}[t]
\includegraphics[width=1\columnwidth]{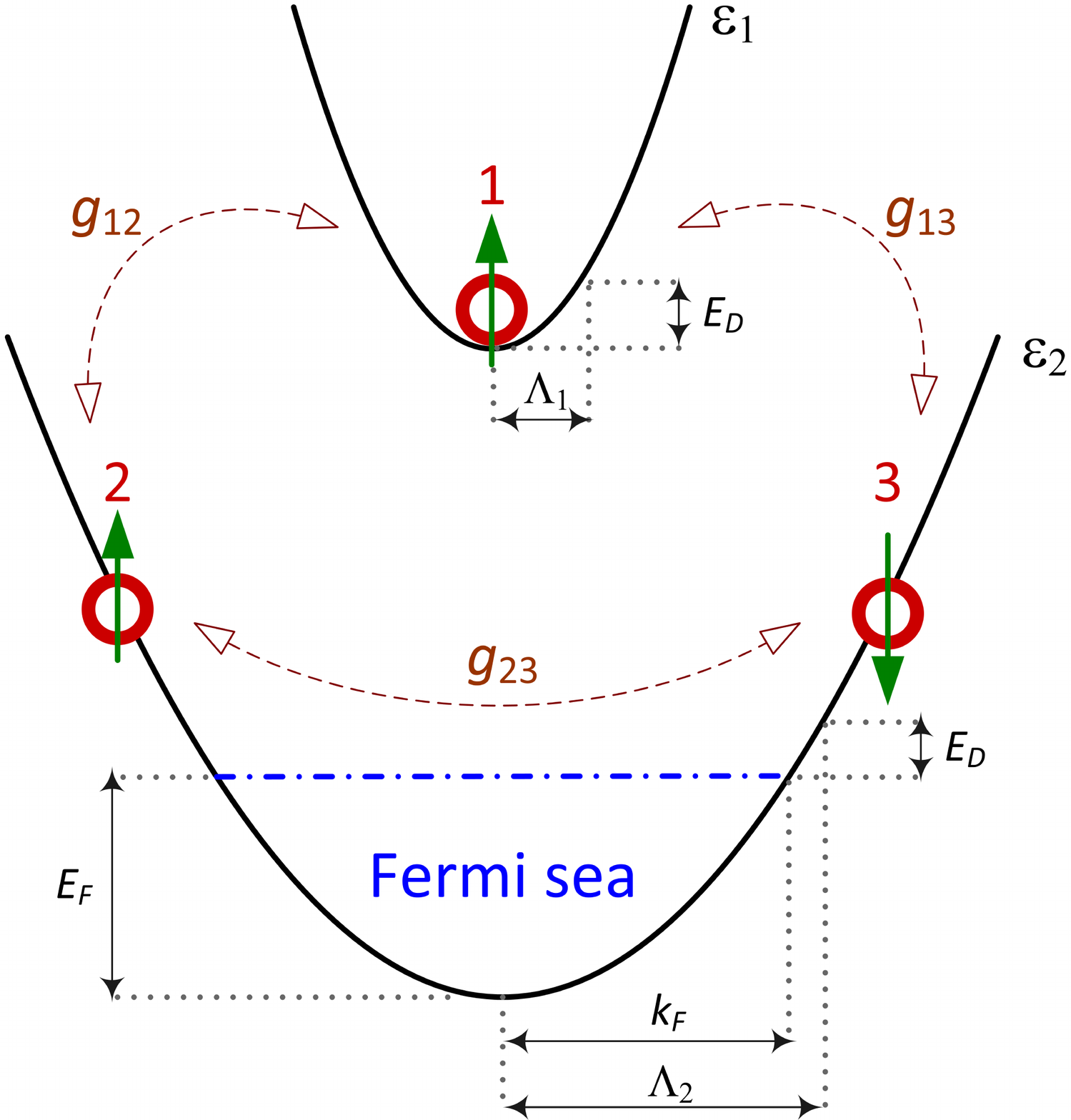}

\caption{The expanded Cooper problem consists of two electrons in a partially
filled band with dispersion $\varepsilon_{2}$ and a third electron
in an empty band with dispersion $\varepsilon_{1}$. The inert Fermi
sea in the lower band has a Fermi energy $E_{F}$ and Fermi momentum
$k_{F}$. The three electrons interact attractively via the two-body
interactions $g_{12}$, $g_{13}$, and $g_{23}$, with $g_{12}=g_{13}$.
These interactions are cut off in momentum space by the cutoffs $\Lambda_{1}$
and $\Lambda_{2}$, whose magnitudes are chosen to fulfill $\varepsilon_{1}(\Lambda_{1})=\varepsilon_{2}(\Lambda_{2})-E_{F}=E_{D}$,
where $E_{D}$ is the Debye energy. }
\label{Fig_1}
\end{figure}

\begin{figure}
\includegraphics[width=1\columnwidth]{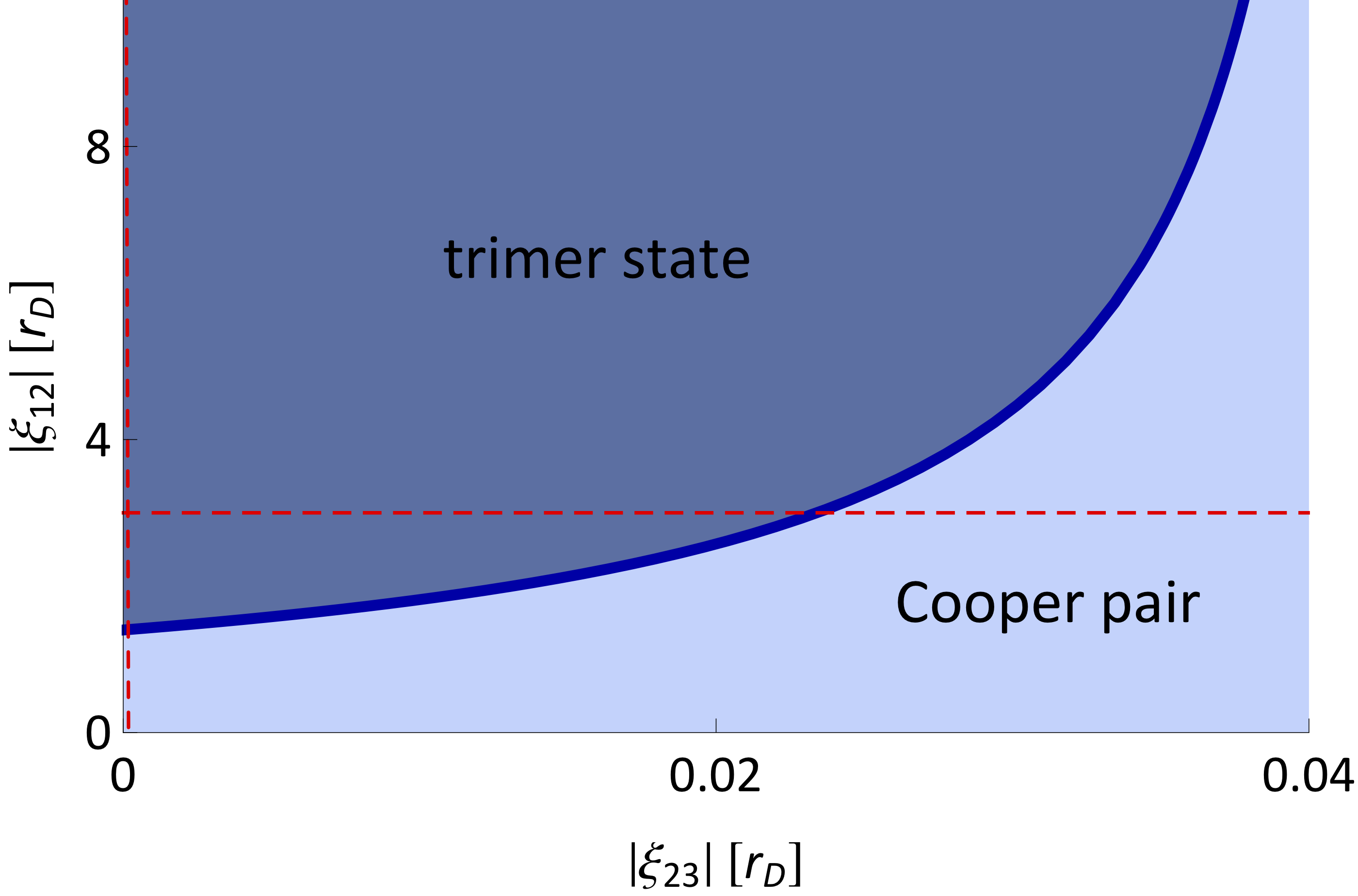}

\caption{Lowest energy state as a function of the interaction parameters $|\xi_{23}|$
and $|\xi_{12}|$, both in units of $r_{D}$. In this example, we
choose $E_{D}/E_{F}=0.02$, $m_{2}/m_{1}=1$, and the cutoffs $\Lambda_{1}$
and $\Lambda_{2}$ according to Eq. (\ref{Debye energy}). For sufficiently
strong attractive inter-band interaction $\xi_{12}$, the trimer state
has lower energy than the Cooper pair. The horizontal dashed red line
is a cut at $|\xi_{12}|=3r_{D}$, see Fig. 3, and the vertical dashed
red line is a cut at $|\xi_{23}|=0$, see Fig. 4.}

\label{Fig_2}
\end{figure}

In this Letter, we expand the Cooper problem by including a third
electron, as depicted in Fig. \ref{Fig_1}. We assume that this additional
electron, labeled `1', is in an otherwise empty band with quadratic
dispersion relation, $\varepsilon_{1}$. Its spin state is arbitrarily
depicted as spin-up. The two electrons `2' and `3' are restricted
to be outside of an inert Fermi sea of momentum $k_{F}$ and of energy
$E_{F}$. For simplicity, the dispersion $\varepsilon_{2}$ of the
lower band is assumed to be quadratic. We propose to realize this
scenario by optically pumping electrons from a lower band into an
unoccupied band, using current technology of pump-probe experiments
\cite{Cavalleri_1,Cavalleri_2,Cavalleri_3,Junichi}. This results
in a low, metastable electron density in the upper band. We assume
that the effective interaction between the electrons is attractive,
following the reasoning of the Cooper problem. We consider the interaction
between two electrons to be a negative constant $g_{ij}$, with $i,j=1,2,3$
and $i\neq j$, for the incoming and outgoing momentum of particle
$i$ smaller than a cutoff $\Lambda_{i}$, and zero otherwise. We
choose the values of $\Lambda_{1}$ and $\Lambda_{2}=\Lambda_{3}$
such that

\begin{equation}
E_{D}=\frac{\hbar^{2}}{2m_{1}}\Lambda_{1}^{2}=\frac{\hbar^{2}}{2m_{2}}(\Lambda_{2}^{2}-k_{F}^{2}),\label{Debye energy}
\end{equation}
where $m_{i}$ is the effective mass of particle $i$ and $E_{D}$
is the Debye energy, see Fig. \ref{Fig_1}. For clarity, we allow
for three different masses $m_{1}$, $m_{2}$, and $m_{3}$, but are
primarily interested in the case $m_{3}=m_{2}$. A similar restriction
will be placed on $g_{12}$ and $g_{13}$, which we choose to be equal
throughout this Letter. For a typical conventional superconductor
we have $E_{D}\ll E_{F}$, implying that $\Lambda_{1}\ll k_{F}$ and
$\Lambda_{2}-k_{F}\ll k_{F}$, which we will use as small parameters
further down. The Fermi velocity is $v_{F}=\hbar k_{F}/m_{2}$. We
define the length scale

\begin{equation}
r_{D}=(\Lambda_{2}-k_{F})^{-1}\approx\hbar\frac{v_{F}}{E_{D}}.\label{length r_D}
\end{equation}
 We note that in the following we consider three-body states with
vanishing total momentum.

The main result of our analysis is shown in Fig. \ref{Fig_2}. We
depict whether the lowest energy state is a Cooper pair state or a
trimer state, as a function of the interaction parameters $\xi_{23}$
and $\xi_{12}=\xi_{13}$, where $\xi_{23}=2\tilde{\mu}/(4\pi\hbar^{2})g_{23}$
and $\xi_{12}=2\mu/(4\pi\hbar^{2})g_{12}$. The reduced masses are
$1/\tilde{\mu}=1/m_{2}+1/m_{3}=2/m_{2}$ and $1/\mu=1/m_{1}+1/m_{2}$.

For $g_{12}=0$ we recover the result of the Cooper problem. For any
value of $g_{23}<0$ the electrons in the lower band form a pair.
As $g_{12}$ is set to a negative nonzero value, we show that the
three electrons form a trimer state beyond a critical value of $g_{12}$,
which increases in absolute magnitude as $|g_{23}|$ increases. This
trimer formation also occurs for vanishing $g_{23}$. From the perspective
of the electrons in the lower band, this can also be considered as
a bound state formation that is induced by a third electron in an
higher band, i.e., a particle-induced bound state. As we show below,
the magnitude of the critical value of $g_{12}$ for trimer formation
is controlled by the ratio of the Fermi velocity, and the mass of
the electron in the upper band and the Debye energy. If the mass of
the electrons in the upper band is heavier, the critical value is
reduced. Similarly, a smaller Fermi velocity and a larger Debye energy
reduces the critical value. We note that for the typical parameter
regime of conventional superconductors we find one trimer state only
\cite{footnote_Efimov_comparison}. However, we also give an example
for a parameter regime in which more than one trimer state exists
below.

\begin{figure}[t]
\includegraphics[width=1\columnwidth]{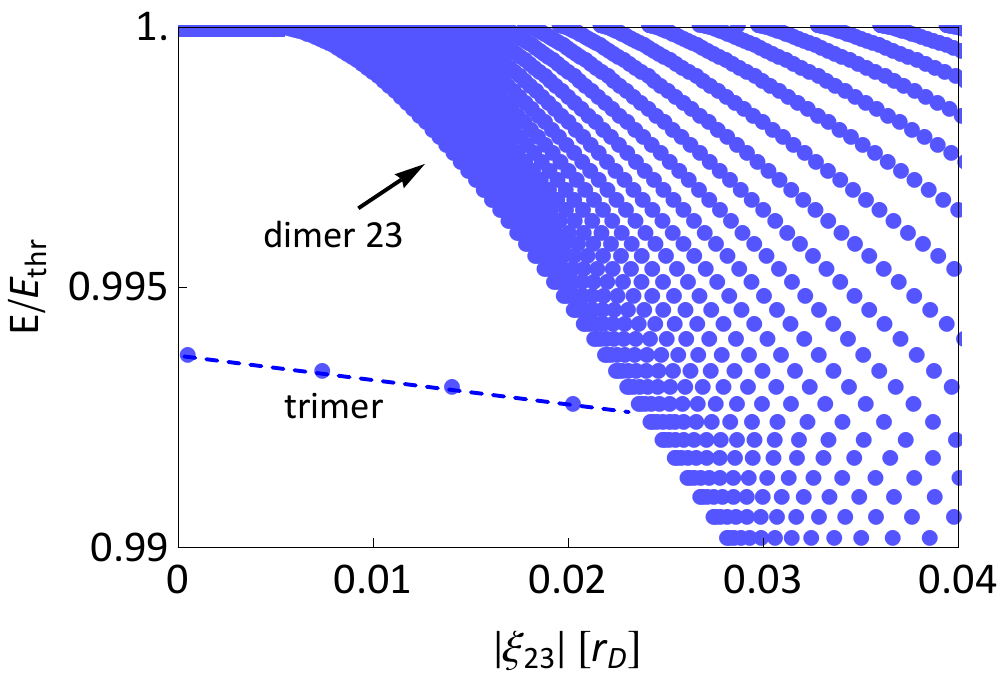}

\caption{Energy $E$ versus the interaction parameter $|\xi_{23}|$, in units
of $r_{D}$, for $E_{D}/E_{F}=0.02$, $m_{2}/m_{1}=1$, and $\xi_{12}=-3r_{D}$,
which corresponds to the horizontal dashed red line in Fig. 2. The
lowest energy state for small values of $|\xi_{23}|$ is a trimer
state. For sufficiently large values of $|\xi_{23}|$ the formation
of Cooper pairs will be dominant over a trimer state.}
\label{Fig_3}
\end{figure}

For a pump-probe experiment, this result implies that a system that
is initially either in a superconducting or a metallic state can be
transformed into a Fermi liquid of electron trimers when electrons
are pumped into a higher band, and the attractive inter-band interaction
is sufficiently strong. 

\begin{figure*}
\includegraphics[scale=0.69]{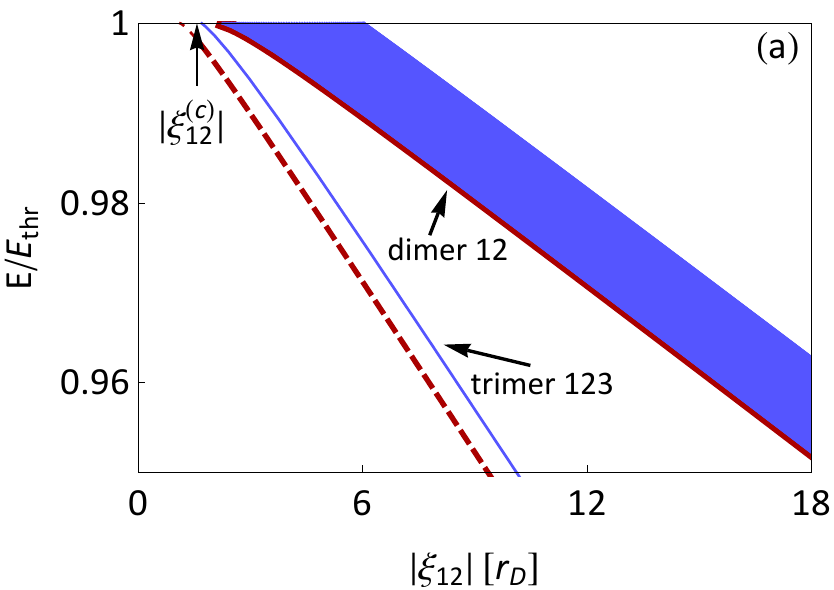}$\quad$\includegraphics[scale=0.65]{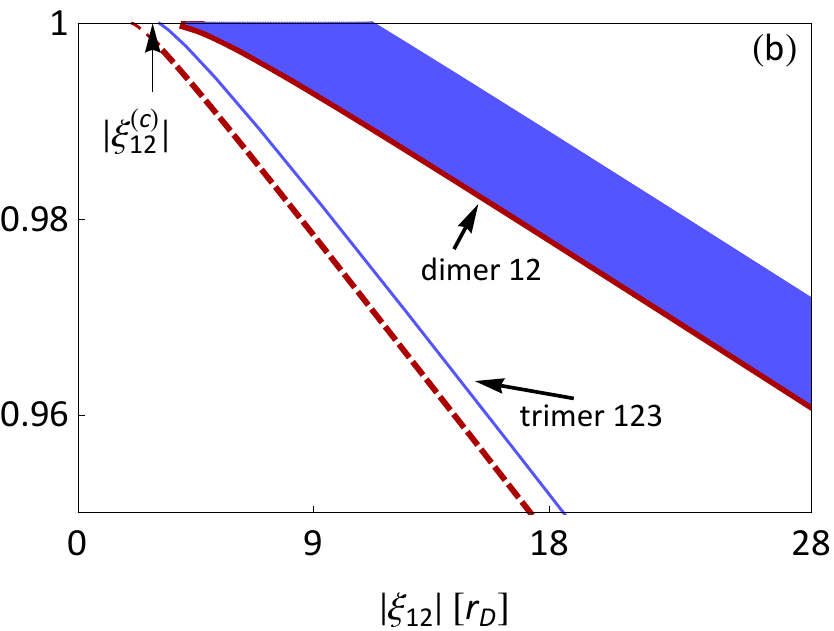}$\quad$\includegraphics[scale=0.65]{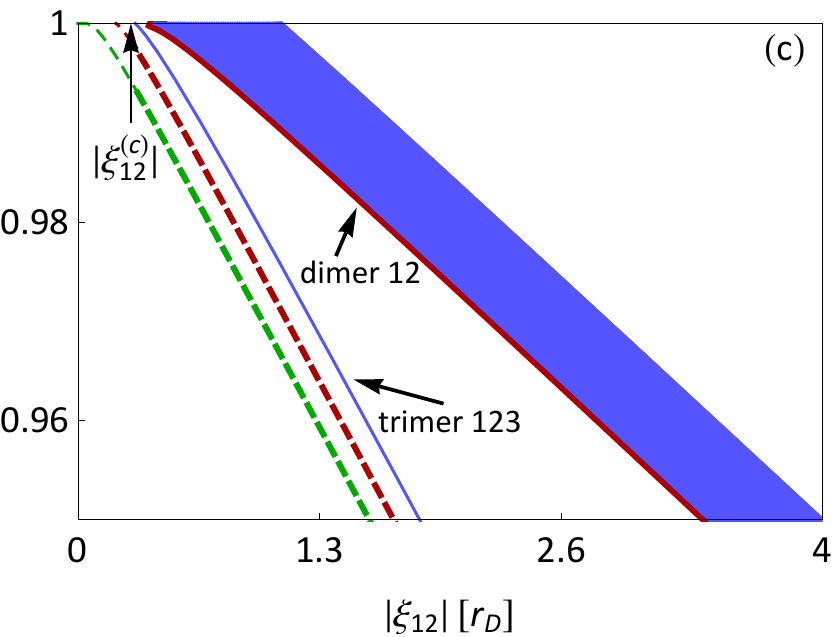}

\caption{The resulting three-body eigenenergies $E$ versus the interaction
parameter $|\xi_{12}|$, in units of $r_{D}$, with $g_{23}=0$ and
$E_{D}/E_{F}=0.02$, for (a) $m_{2}/m_{1}=1$, which corresponds to
the vertical dashed red line in Fig. 2, (b) $m_{2}/m_{1}=10$, and
(c) $m_{2}/m_{1}=1/10$. In each panel the single blue curve is the
numerical solution of the lowest energy trimer state. As $|\xi_{12}|$
increases, the first pair (dimer 12) appears as the lowest energy
state of a two-body bound state continuum depicted by dense blue curves.
The vertical arrow locates the critical value of the inter-band interaction
parameter given by Eq. (\ref{g_12 critical}), which is in good agreement
with the onset of the numerical result. The dashed red curve shows
the analytical approximation for the trimer state, Eq. (\ref{first approximation}),
derived in App. D. It represents a good approximation for the asymptotic
of the single blue curve, which is depicted by thicker red. The solid
red curve shows the analytical approximation for the lowest energy
pair (dimer 12), Eq. (\ref{approximation for dimer}), derived in
App. D, which is in good agreement with numerics. In panel (c) the
green dashed curve is the second analytical approximation (\ref{energy for high m1}).
The thicker green curve represents a good approximation for the asymptotic
of the single blue curve as long as $m_{1}\gg m_{2}$. }
\label{Fig_4}
\end{figure*}

To determine the bound states of the three-electron system we consider
the Schr{\"o}dinger equation in momentum space:

\begin{equation}
\left(\frac{\hbar^{2}k_{1}^{2}}{2m_{1}}+\frac{\hbar^{2}k_{2}^{2}}{2m_{2}}+\frac{\hbar^{2}k_{3}^{2}}{2m_{3}}+\hat{U}_{12}+\hat{U}_{13}+\hat{U}_{23}-E\right)\psi=0,\label{Schrodinger Eq}
\end{equation}
where $\mathbf{k}_{1}$, $\mathbf{k}_{2}$, $\mathbf{k}_{3}$ are
the electron momenta, $E$ is the energy \cite{units}, and $\psi=\psi(\mathbf{k}_{1},\mathbf{k}_{2},\mathbf{k}_{3})$
is the wave function. The interaction $\hat{U}_{ij}$ between two
electrons `$i$' and `$j$' is

\begin{equation}
\hat{U}_{ij}=g_{ij}\theta_{\Lambda_{i}}(\mathbf{k}_{i})\theta_{\Lambda_{j}}(\mathbf{k}_{j})\int\frac{d^{3}\mathbf{q}}{(2\pi)^{3}}\theta_{\Lambda_{i}}(\mathbf{k}_{i}-\mathbf{q})\theta_{\Lambda_{j}}(\mathbf{k}_{j}+\mathbf{q}),\label{U_ij}
\end{equation}
where $g_{ij}<0$ and $\mathbf{q}$ denotes the momentum transfer
\cite{momentum transfer}. The resulting operators $\hat{U}_{ij}\psi$
are represented in App. A. The cutoff function $\theta_{a,b}(\mathbf{k})$
is defined as

\begin{equation}
\theta_{a,b}(\mathbf{k})=\begin{cases}
1 & \text{for }a\leqslant|\mathbf{k}|\leqslant b,\\
0 & \mathrm{\text{otherwise,}}
\end{cases}\label{cutoff function}
\end{equation}
for two real numbers $0\leqslant a<b$, and $\theta_{b}(\mathbf{k})\equiv\theta_{0,b}(\mathbf{k})$.
The inert Fermi sea demands the constraints $k_{2}>k_{F}$ and $k_{3}>k_{F}$
on the momenta of electrons `2' and `3', respectively. We consider
a singlet state for the electrons `2' and `3' in the following.
This system is separable, as shown in App. A, which results in a system
of two coupled integral equations:

\begin{widetext}

\begin{equation}
\left[\frac{1}{g_{12}}+\int\frac{d^{3}\mathbf{p}_{3}}{(2\pi)^{3}}K_{1}(\mathbf{k}_{2},\mathbf{p}_{3};E)\right]F_{2}(\mathbf{k}_{2})=-\theta_{k_{F},\Lambda_{2}}(\mathbf{k}_{2})\left[\int\frac{d^{3}\mathbf{p}_{3}}{(2\pi)^{3}}K_{1}(\mathbf{k}_{2},\mathbf{p}_{3};E)F_{2}(\mathbf{p}_{3})+\int\frac{d^{3}\mathbf{p}_{1}}{(2\pi)^{3}}K_{2}(\mathbf{k}_{2},\mathbf{p}_{1};E)F_{1}(\mathbf{p}_{1})\right],\label{Main Eq 1}
\end{equation}

\begin{equation}
\left[\frac{1}{g_{23}}+\int\frac{d^{3}\mathbf{p}_{3}}{(2\pi)^{3}}K_{3}(\mathbf{k}_{1},\mathbf{p}_{3};E)\right]F_{1}(\mathbf{k}_{1})=-2\theta_{\Lambda_{1}}(\mathbf{k}_{1})\int\frac{d^{3}\mathbf{p}_{3}}{(2\pi)^{3}}K_{3}(\mathbf{k}_{1},\mathbf{p}_{3};E)F_{2}(\mathbf{p}_{3}),\label{Main Eq 2}
\end{equation}

\end{widetext}where $\mathbf{p}_{i}=\mathbf{k}_{i}-\mathbf{q}$,
for $i=1,2$, and $\mathbf{p}_{3}=\mathbf{k}_{3}+\mathbf{q}$. The
three integral kernels $K_{1}$, $K_{2}$, $K_{3}$ and the three
functions $F_{1}$, $F_{2}$, $F_{3}$ are derived in App. A. Due
to the singlet symmetry for electrons `2' and `3' we consider
$F_{2}=F_{3}$. We assume $F_{i}(\mathbf{k})=F_{i}(k)$, implying
\emph{s}-wave symmetry of the state.

To determine whether the lowest energy state is a two-body or a three-body
bound state, we use the small parameter $E_{D}/E_{F}$ to approximate
the full integral equation with an equation that relates $\xi_{23}$,
$\xi_{12}$, and $E$, as described on App. B. We solve this equation
numerically for energies near the threshold energy $E_{\mathrm{thr}}=2E_{F}=\hbar^{2}k_{F}^{2}/m_{2}$
which results in Fig. \ref{Fig_2}, depicting a region where electrons
`2' and `3' form a Cooper pair, and a second region where the
three electrons form a trimer state. The intra-band electrons `2'
and `3' can form a Cooper pair for any attractive interaction $\xi_{23}$.
The trimer state is only formed when the inter-band interaction $\xi_{12}$
is sufficiently strong. Trimer states of zero total momentum appear
as discrete energy levels, whereas dimer states appear as continuum
of states.

Next, we solve the full Eqs. (\ref{Main Eq 1}) and (\ref{Main Eq 2})
numerically. For that, we reduce the three dimensional integrals over
momentum to one dimensional integrals over the absolute value of each
momentum. We approximate the integrals by a sum over discrete values
according to the Gauss-Legendre quadrature rule \cite{Gaussian_quadrature_1,Gaussian_quadrature_2,Gaussian_quadrature_3}.
The continuous functions $F_{1}$ and $F_{2}$ are evaluated at these
discrete momentum values. We therefore approximate Eqs. (\ref{Main Eq 1})
and (\ref{Main Eq 2}) with a discrete eigenvalue problem, see App.
C. 

In Fig. \ref{Fig_3} we show the resulting eigenenergies $E$ below
the threshold energy $E_{\mathrm{thr}}$, for $E_{D}/E_{F}=0.02$
and $\xi_{12}=-3r_{D}$. For small values of $|\xi_{23}|$ the lowest
energy state is a trimer state which appears as a single line of solutions.
For larger values the lowest energy state is a Cooper pair which appears
as the lowest energy state of a two-body bound state continuum.

To estimate the critical value of $g_{12}$ for vanishing $g_{23}$
analytically, we recall that $F_{1}\propto g_{23}$ and $F_{2}\propto g_{12}$,
cf. App. A. With this, Eqs. (\ref{Main Eq 1}) and (\ref{Main Eq 2})
reduce to

\begin{align}
\frac{4\pi\hbar^{2}}{2\mu g_{12}}+\frac{\tau}{\frac{\mu}{m_{1}}\pi k_{F}}\int_{k_{F}}^{\Lambda_{2}}dp_{3}\,p_{3}\times\qquad\qquad\qquad\qquad\quad\nonumber \\
\times\ln\left[\frac{(1-\frac{\mu}{m_{1}})p_{3}^{2}+(1-\frac{\mu}{m_{1}})k_{F}^{2}+\frac{\mu}{m_{1}}\Lambda_{1}^{2}-\frac{2\mu}{\hbar^{2}}E}{p_{3}^{2}-\frac{2\mu}{m_{1}}k_{F}p_{3}+k_{F}^{2}-\frac{2\mu}{\hbar^{2}}E}\right] & \approx0,\label{Eq for g23 zero}
\end{align}
where $\tau=1$ describes the system of three electrons and $\tau=1/2$
corresponds to a system of two electrons `1' and `2' (or `3'),
see App. D. We evaluate the integral, which is done in App. B and
D, and expand to lowest order in $E_{D}/E_{F}$. We solve for the
interaction parameter $g_{12}$, then choose the threshold condition
$E=E_{\mathrm{thr}}$, which finally gives the critical value

\begin{figure}
\includegraphics{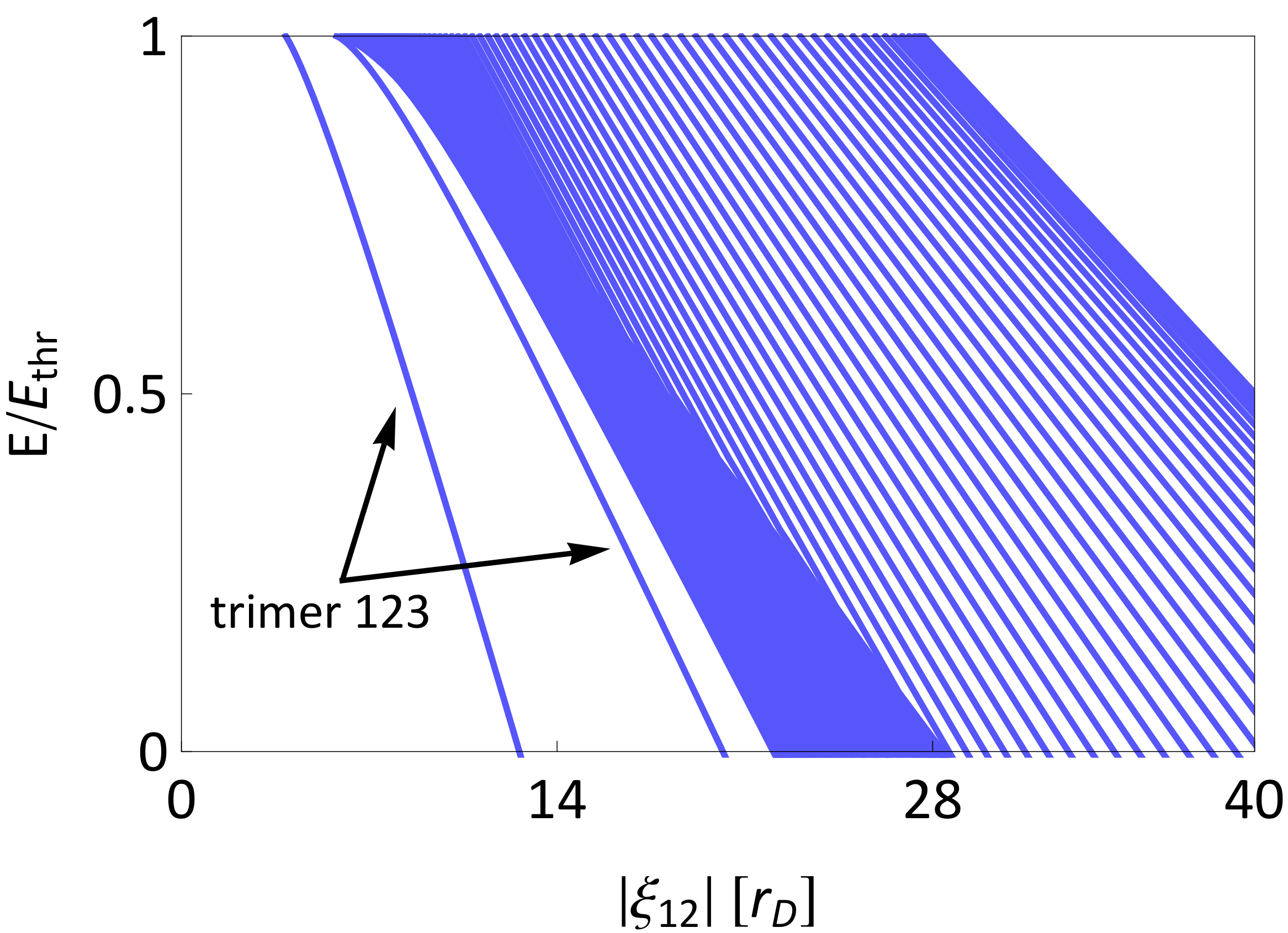}

\caption{The resulting three-body eigenenergies $E$ versus the interaction
parameter $|\xi_{12}|$, in units of $r_{D}$, with $g_{23}=0$ and
$E_{D}\sim E_{F}$, for $m_{2}/m_{1}=10$. Here, we see the formation
of two trimer states that are shown by two single blue curves. The
dense curves show the two-body bound state continuum. When $m_{2}\gg m_{1}$
we can see the formation of more than one trimer states, however,
the number of the states will remain finite.}

\label{Fig_5}
\end{figure}

\begin{equation}
|g_{12}^{(\mathrm{c})}|\sim\frac{2\pi^{2}\hbar^{2}}{m_{1}}r_{D}\approx\frac{2\pi^{2}\hbar^{3}}{m_{1}}\frac{v_{F}}{E_{D}},\label{g_12 critical}
\end{equation}
where $v_{F}=\hbar k_{F}/m_{2}$. This shows that a lower value of
the critical interaction strength is achieved for heavier electrons
in the upper band, for a smaller Fermi velocity, and a higher Debye
energy.

An approximate analytical solution of Eq. (\ref{Eq for g23 zero}),
which describes both the trimer state and the lowest energy two-body
bound state, is derived in App. D, based on Eqs. (\ref{first approximation})
and (\ref{approximation for dimer}), respectively. These solutions
are depicted as the dashed and the continuous line in Figs. \ref{Fig_4}
(a)-(c). For $m_{1}\gg m_{2}$ and $E_{D}\ll E_{F}$ we have $|\xi_{12}^{(\mathrm{c})}|=2\mu/(4\pi\hbar^{2})|g_{12}^{(\mathrm{c})}|\rightarrow0$.
In this case, a second analytical approximate solution of Eq. (\ref{Eq for g23 zero})
can be represented by:

\begin{align}
E & \approx E_{\mathrm{thr}}+\frac{E_{D}}{1-\exp\left(\pi\frac{\mu}{m_{1}}\frac{r_{D}}{|\xi_{12}|}\right)},\label{energy for high m1}
\end{align}
see App. D. This approximation is shown as a green dashed line in
Fig. \ref{Fig_4} (c). For larger $m_{1}/m_{2}$ Eq. (\ref{energy for high m1})
becomes a better approximation. We compare these analytical results
with the numerical results for different parameter sets in Fig. \ref{Fig_4}.
In \ref{Fig_4} (a) we use $m_{2}=m_{1}$ and $E_{D}/E_{F}=0.02$.
In 4 (b) we use $m_{2}/m_{1}=10$ and $E_{D}/E_{F}=0.02$. In 4 (c)
we use $m_{2}/m_{1}=1/10$ and $E_{D}/E_{F}=0.02$. We observe the
formation of a trimer state, that is well approximated by the analytical
approximation. In contrast to the Cooper problem where a two-body
bound state originates at a vanishing coupling constant \cite{Cooper_seminal_paper,Tinkham,Baym},
here, a trimer state emerges at the critical value $\xi_{12}^{(\mathrm{c})}$.

Finally, for $m_{2}\gg m_{1}$ and large values of the Debye energy
that are comparable with Fermi energy, $E_{D}\sim E_{F}$, we observe
the formation of more than one trimer state \cite{footnote_Efimov_comparison}.
Figure \ref{Fig_5} shows the formation of two trimer states for $m_{2}/m_{1}=10$.
Physical systems in this regime are superconductors like fullerides
\cite{fullerides} or magnesium diboride \cite{MgB2}. 

In conclusion, we have demonstrated the formation of a trimer state
of electrons in a conventional superconductor, in which an additional
electron occupies a higher band. We show this by expanding the Cooper
problem of two attractively interacting electrons by adding an additional
electron that also interacts attractively with the other electrons.
The trimer formation sets in beyond a critical inter-band interaction
strength, for which we give an analytical estimate. This demonstrates
an instability of the BCS state. We propose to realize this scenario
by optically pumping electrons to otherwise single empty band. Out
of the initial superconducting or metallic state, a transient state
of a Fermi liquid of electron trimers can be formed. We also emphasize
that the analogue of the Cooper problem can also be formulated for
orders such as antiferromagnetism or charge-density wave orders. Here,
we have a two-body problem of an electron and a hole. We emphasize
that the analysis of this paper can be extended to any order that
is described by a two-Fermi order parameter, and predicts three-fermion
bound states for all these orders for the corresponding parameter
regimes.

\emph{Acknowledgments}.\textendash{} AS and LM would like to acknowledge
support from the Deutsche Forschungsgemeinschaft through SFB 925.
PN acknowledges support from the RIKEN Incentive Research Projects.

\section*{appendices}

\numberwithin{equation}{subsection}

\subsection{{\normalsize{}Derivation of the system of two coupled integral equations
(\ref{Main Eq 1}) and (\ref{Main Eq 2}) }}

Applying the interaction operators $\hat{U}_{ij}$, given by Eq. (\ref{U_ij}),
on the wave function $\psi=\psi(\mathbf{k}_{1},\mathbf{k}_{2},\mathbf{k}_{3})$
governing the three-body system in momentum space, we write the Schr{\"o}dinger
equation (\ref{Schrodinger Eq}) as follows:

\begin{align}
\left(\frac{\hbar^{2}k_{1}^{2}}{2m_{1}}+\frac{\hbar^{2}k_{2}^{2}}{2m_{2}}+\frac{\hbar^{2}k_{3}^{2}}{2m_{3}}-E\right)\psi=-\left(\hat{U}_{12}+\hat{U}_{13}+\hat{U}_{23}\right)\psi\nonumber \\
=\;\;-g_{12}\theta_{\Lambda_{1}}(\mathbf{k}_{1})\theta_{\Lambda_{2}}(\mathbf{k}_{2})\times\nonumber \\
\times\int\frac{d^{3}\mathbf{q}}{(2\pi)^{3}}\theta_{\Lambda_{1}}(\mathbf{k}_{1}-\mathbf{q})\theta_{\Lambda_{2}}(\mathbf{k}_{2}+\mathbf{q})\psi(\mathbf{k}_{1}-\mathbf{q},\mathbf{k}_{2}+\mathbf{q},\mathbf{k}_{3})\nonumber \\
-g_{13}\theta_{\Lambda_{1}}(\mathbf{k}_{1})\theta_{\Lambda_{3}}(\mathbf{k}_{3})\times\nonumber \\
\times\int\frac{d^{3}\mathbf{q}}{(2\pi)^{3}}\theta_{\Lambda_{1}}(\mathbf{k}_{1}-\mathbf{q})\theta_{\Lambda_{3}}(\mathbf{k}_{3}+\mathbf{q})\psi(\mathbf{k}_{1}-\mathbf{q},\mathbf{k}_{2},\mathbf{k}_{3}+\mathbf{q})\nonumber \\
-g_{23}\theta_{\Lambda_{2}}(\mathbf{k}_{2})\theta_{\Lambda_{3}}(\mathbf{k}_{3})\times\nonumber \\
\times\int\frac{d^{3}\mathbf{q}}{(2\pi)^{3}}\theta_{\Lambda_{2}}(\mathbf{k}_{2}-\mathbf{q})\theta_{\Lambda_{3}}(\mathbf{k}_{3}+\mathbf{q})\psi(\mathbf{k}_{1},\mathbf{k}_{2}-\mathbf{q},\mathbf{k}_{3}+\mathbf{q}),\label{Integral Eq 1}
\end{align}
where $\theta_{b}(\mathbf{k})\equiv\theta_{0,b}(\mathbf{k})$ and
$\theta_{a,b}(\mathbf{k})$ is defined by Eq. (\ref{cutoff function}).
Introducing three variables $\tilde{\mathbf{p}}_{i}\equiv\mathbf{q}+\mathbf{k}_{i}$,
with $i=1,2,3$, and assuming the zero total momentum of the system,
$\psi(\mathbf{k}_{1},\mathbf{k}_{2},\mathbf{k}_{3})=\psi(\mathbf{k}_{2},\mathbf{k}_{3})\delta^{(3)}(\mathbf{k}_{1},\mathbf{k}_{2},\mathbf{k}_{3})$,
where $\delta^{(3)}$ is the three-dimensional Dirac delta function,
we rewrite Eq. (\ref{Integral Eq 1}) in the following form:

\begin{align}
\left[\frac{\hbar^{2}(\mathbf{k}_{2}+\mathbf{k}_{3})^{2}}{2m_{1}}+\frac{\hbar^{2}k_{2}^{2}}{2m_{2}}+\frac{\hbar^{2}k_{3}^{2}}{2m_{3}}-E\right]\psi(\mathbf{k}_{2},\mathbf{k}_{3})\qquad\qquad\nonumber \\
=-\theta_{\Lambda_{1}}(-\mathbf{k}_{2}-\mathbf{k}_{3})\theta_{\Lambda_{2}}(\mathbf{k}_{2})F_{3}(\mathbf{k}_{3})-\theta_{\Lambda_{1}}(-\mathbf{k}_{2}-\mathbf{k}_{3})\times\nonumber \\
\times\theta_{\Lambda_{2}}(\mathbf{k}_{3})F_{2}(\mathbf{k}_{2})-\theta_{\Lambda_{2}}(\mathbf{k}_{2})\theta_{\Lambda_{3}}(\mathbf{k}_{3})F_{1}(-\mathbf{k}_{2}-\mathbf{k}_{3}).\label{integral Eq. 2}
\end{align}
Here, the functions $F_{1}$, $F_{2}$, and $F_{3}$ are defined by
\begin{align}
F_{1}(\mathbf{k}_{1}) & =g_{23}\int\frac{d^{3}\tilde{\mathbf{p}}_{3}}{(2\pi)^{3}}\theta_{\Lambda_{2}}(-\mathbf{k}_{1}-\tilde{\mathbf{p}}_{3})\theta_{\Lambda_{3}}(\tilde{\mathbf{p}}_{3})\times\nonumber \\
 & \qquad\qquad\qquad\qquad\qquad\times\psi(-\mathbf{k}_{1}-\tilde{\mathbf{p}}_{3},\tilde{\mathbf{p}}_{3}),\label{F1_1}\\
F_{2}(\mathbf{k}_{2}) & =g_{13}\int\frac{d^{3}\tilde{\mathbf{p}}_{3}}{(2\pi)^{3}}\theta_{\Lambda_{1}}(-\mathbf{k}_{2}-\tilde{\mathbf{p}}_{3})\theta_{\Lambda_{3}}(\tilde{\mathbf{p}}_{3})\psi(\mathbf{k}_{2},\tilde{\mathbf{p}}_{3}),\label{F2 _1}\\
F_{3}(\mathbf{k}_{3}) & =g_{12}\int\frac{d^{3}\tilde{\mathbf{p}}_{2}}{(2\pi)^{3}}\theta_{\Lambda_{1}}(-\mathbf{k}_{3}-\tilde{\mathbf{p}}_{2})\theta_{\Lambda_{2}}(\tilde{\mathbf{p}}_{2})\psi(\tilde{\mathbf{p}}_{2},\mathbf{k}_{3}).\label{F3_1}
\end{align}
Equation (\ref{integral Eq. 2}) provides now an ansatz for the wave
function $\psi(\mathbf{k}_{2},\mathbf{k}_{3})$:

\begin{widetext}

\begin{equation}
\psi(\mathbf{k}_{2},\mathbf{k}_{3})=-\frac{\theta_{\Lambda_{1}}(-\mathbf{k}_{2}-\mathbf{k}_{3})\theta_{\Lambda_{2}}(\mathbf{k}_{2})F_{3}(\mathbf{k}_{3})+\theta_{\Lambda_{1}}(-\mathbf{k}_{2}-\mathbf{k}_{3})\theta_{\Lambda_{3}}(\mathbf{k}_{3})F_{2}(\mathbf{k}_{2})+\theta_{\Lambda_{2}}(\mathbf{k}_{2})\theta_{\Lambda_{3}}(\mathbf{k}_{3})F_{1}(-\mathbf{k}_{2}-\mathbf{k}_{3})}{\frac{\hbar^{2}(\mathbf{k}_{2}+\mathbf{k}_{3})^{2}}{2m_{1}}+\frac{\hbar^{2}k_{2}^{2}}{2m_{2}}+\frac{\hbar^{2}k_{3}^{2}}{2m_{3}}-E}.\label{wavefunction ansatz}
\end{equation}
\end{widetext}If the electrons `2' and `3' are in a spin singlet
state and $g_{12}=g_{13}$, then $F_{3}=F_{2}$. We also assume $m_{3}=m_{2}$
and take into account the Fermi sea condition by $k_{2},k_{3}>k_{F}$.
Introducing the variables $\mathbf{p}_{1}\equiv-\mathbf{k}_{2}-\tilde{\mathbf{p}}_{3}$
and $\mathbf{p}_{2}\equiv-\mathbf{k}_{1}-\tilde{\mathbf{p}}_{3}$,
with $\mathbf{p}_{3}\equiv\tilde{\mathbf{p}}_{3}$, the functions
$F_{1}$ and $F_{2}$ now read:

\begin{align}
F_{1}(\mathbf{k}_{1}) & =g_{23}\int\frac{d^{3}\mathbf{p}_{3}}{(2\pi)^{3}}\theta_{k_{F},\Lambda_{2}}(-\mathbf{k}_{1}-\mathbf{p}_{3})\theta_{k_{F},\Lambda_{2}}(\mathbf{p}_{3})\times\nonumber \\
 & \qquad\qquad\qquad\qquad\qquad\times\psi(-\mathbf{k}_{1}-\mathbf{p}_{3},\mathbf{p}_{3}),\label{F1 _2}\\
F_{2}(\mathbf{k}_{2}) & =g_{12}\int\frac{d^{3}\mathbf{p}_{3}}{(2\pi)^{3}}\theta_{\Lambda_{1}}(-\mathbf{k}_{2}-\mathbf{p}_{3})\theta_{k_{F},\Lambda_{2}}(\mathbf{p}_{3})\times\nonumber \\
 & \qquad\qquad\qquad\qquad\qquad\times\psi(\mathbf{k}_{2},\mathbf{p}_{3}).\label{F2 _2}
\end{align}
We insert the ansatz (\ref{wavefunction ansatz}) into Eqs. (\ref{F1 _2})
and (\ref{F2 _2}), and arrive at a system of two coupled integral
equations (\ref{Main Eq 1}) and (\ref{Main Eq 2}), where the three
kernels $K_{1}$, $K_{2}$, and $K_{3}$ are given by:
\begin{align}
K_{1}(\mathbf{k}_{2},\mathbf{p}_{3};E) & =\frac{\theta_{\Lambda_{1}}(-\mathbf{k}_{2}-\mathbf{p}_{3})\theta_{k_{F},\Lambda_{2}}(\mathbf{p}_{3})}{\frac{\hbar^{2}(\mathbf{k}_{2}+\mathbf{p}_{3})^{2}}{2m_{1}}+\frac{\hbar^{2}k_{2}^{2}}{2m_{2}}+\frac{\hbar^{2}p_{3}^{2}}{2m_{2}}-E},\label{K1}\\
K_{2}(\mathbf{k}_{2},\mathbf{p}_{1};E) & =\frac{\theta_{\Lambda_{1}}(\mathbf{p}_{1})\theta_{k_{F},\Lambda_{2}}(-\mathbf{p}_{1}-\mathbf{k}_{2})}{\frac{\hbar^{2}p_{1}^{2}}{2m_{1}}+\frac{\hbar^{2}k_{2}^{2}}{2m_{2}}+\frac{\hbar^{2}(\mathbf{p}_{1}+\mathbf{k}_{2})^{2}}{2m_{2}}-E},\label{K2}\\
K_{3}(\mathbf{k}_{1},\mathbf{p}_{3};E) & =\frac{\theta_{k_{F},\Lambda_{2}}(-\mathbf{k}_{1}-\mathbf{p}_{3})\theta_{k_{F},\Lambda_{2}}(\mathbf{p}_{3})}{\frac{\hbar^{2}k_{1}^{2}}{2m_{1}}+\frac{\hbar^{2}(\mathbf{k}_{1}+\mathbf{p}_{3})^{2}}{2m_{2}}+\frac{\hbar^{2}p_{3}^{2}}{2m_{2}}-E}.\label{K3}
\end{align}

\subsection{{\normalsize{}Overall behavior of the three-electron system}}

As mentioned, we choose the values of $\Lambda_{1}$ and $\Lambda_{2}$
according to relation (\ref{Debye energy}). For a typical conventional
superconductor we have $E_{D}\ll E_{F}$, implying that $\Lambda_{1}\ll k_{F}$
and $\Lambda_{2}-k_{F}\ll k_{F}$. We recall that $0<k_{1}<\Lambda_{1}$
and $k_{F}<k_{2}<\Lambda_{2}$. We thus make a first approximation
such that $k_{1}\sim0$ and $k_{2}\sim k_{F}$. In addition, because
the integral variable $p_{3}$ is varying within the interval $(k_{F},\Lambda_{2})$
and $\Lambda_{2}-k_{F}\ll k_{F}$, we make a second approximation
in this interval and assume that the two functions $F_{2}(k_{2})$
and $F_{2}(p_{3})$ both remain constant, $F_{2}(k_{F})$. Thus, we
rewrite the system of Eqs. (\ref{Main Eq 1}) and (\ref{Main Eq 2})
as follows:

\begin{equation}
\begin{cases}
\Omega_{1}(E,g_{12};\tau)F_{2}(k_{F})+\Omega_{2}(E)F_{1}(0)\approx0,\\
\Omega_{3}(E)F_{2}(k_{F})+\Omega_{4}(E,g_{23})F_{1}(0)\approx0.
\end{cases}\label{rewritten sys of integral Eqs}
\end{equation}
We calculate $\Omega_{1}(E,g_{12};\tau)$ to be:

\begin{multline}
\Omega_{1}(E,g_{12};\tau)=\frac{4\pi\hbar^{2}}{2\mu g_{12}}+\frac{\tau}{\frac{\mu}{m_{1}}\pi k_{F}}\int_{k_{F}}^{\Lambda_{2}}dp_{3}\,p_{3}\times\\
\times\ln\left[\frac{(1-\frac{\mu}{m_{1}})p_{3}^{2}+(1-\frac{\mu}{m_{1}})k_{F}^{2}+\frac{\mu}{m_{1}}\Lambda_{1}^{2}-\frac{2\mu}{\hbar^{2}}E}{p_{3}^{2}-\frac{2\mu}{m_{1}}k_{F}p_{3}+k_{F}^{2}-\frac{2\mu}{\hbar^{2}}E}\right]\\
=\frac{4\pi\hbar^{2}}{2\mu g_{12}}+\frac{\tau(\Lambda_{2}-k_{F})}{\pi}+\frac{2\tau\sqrt{\eta(E)}}{\pi}\times\\
\times\left[\arctan\left(\frac{\frac{\mu}{m_{1}}k_{F}-\Lambda_{2}}{\sqrt{\eta(E)}}\right)-\arctan\left(\frac{(\frac{\mu}{m_{1}}-1)k_{F}}{\sqrt{\eta(E)}}\right)\right]\\
+\frac{\tau\Lambda_{2}^{2}}{\frac{2\mu}{m_{1}}\pi k_{F}}\ln\left[\frac{-\rho(E)+(1-\frac{\mu}{m_{1}})\Lambda_{2}^{2}}{\chi(E)}\right]\\
+\frac{\tau\rho(E)}{\frac{2\mu}{m_{1}}(\frac{\mu}{m_{1}}-1)\pi k_{F}}\ln\left[\frac{\rho(E)+(\frac{\mu}{m_{1}}-1)\Lambda_{2}^{2}}{\rho(E)+(\frac{\mu}{m_{1}}-1)k_{F}^{2}}\right]\\
-\frac{\tau[\eta(E)-(\frac{\mu}{m_{1}})^{2}k_{F}^{2}]}{\frac{2\mu}{m_{1}}\pi k_{F}}\ln\left[\frac{\chi(E)}{-\rho(E)+(1-\frac{\mu}{m_{1}})k_{F}^{2}-\frac{\mu}{m_{1}}\Lambda_{1}^{2}}\right]\\
-\frac{\tau k_{F}}{\frac{2\mu}{m_{1}}\pi}\ln\left[\frac{\rho(E)+(\frac{\mu}{m_{1}}-1)k_{F}^{2}}{\rho(E)+(\frac{\mu}{m_{1}}-1)k_{F}^{2}-\frac{\mu}{m_{1}}\Lambda_{1}^{2}}\right],\label{omega_1}
\end{multline}
where $\eta(E)=[1-(\frac{\mu}{m_{1}})^{2}]k_{F}^{2}-\frac{2\mu}{\hbar^{2}}E$,
$\rho(E)=(\frac{\mu}{m_{1}}-1)k_{F}^{2}-\frac{\mu}{m_{1}}\Lambda_{1}^{2}+\frac{2\mu}{\hbar^{2}}E$,
and $\chi(E)=k_{F}^{2}-\frac{2\mu}{m_{1}}k_{F}\Lambda_{2}+\Lambda_{2}^{2}-\frac{2\mu}{\hbar^{2}}E$.
We also obtain:

\begin{align}
\Omega_{2}(E) & =\frac{1}{\frac{2\mu}{m_{2}}\pi k_{F}}\int_{0}^{\Lambda_{1}}dp_{1}\,p_{1}\times\nonumber \\
 & \qquad\;\times\ln\left(\frac{p_{1}^{2}+\frac{2\mu}{m_{2}}k_{F}p_{1}+k_{F}^{2}-\frac{2\mu}{\hbar^{2}}E}{p_{1}^{2}-\frac{2\mu}{m_{2}}k_{F}p_{1}+k_{F}^{2}-\frac{2\mu}{\hbar^{2}}E}\right)\nonumber \\
 & \approx\frac{2}{3\pi}\frac{\Lambda_{1}^{3}}{k_{F}^{2}-\frac{2\mu}{\hbar^{2}}E},\label{omega_2}
\end{align}

\begin{align}
\Omega_{3}(E) & =\frac{2}{\frac{2\tilde{\mu}}{m_{2}}\pi k_{F}}\int_{k_{F}}^{\Lambda_{2}}dp_{3}\,\frac{p_{3}}{k_{1}}\times\nonumber \\
 & \qquad\;\times\ln\left(\frac{\frac{2\tilde{\mu}}{m_{2}}p_{3}^{2}+\frac{2\tilde{\mu}}{m_{2}}k_{1}p_{3}+\frac{\tilde{\mu}}{\mu}k_{F}^{2}-\frac{\tilde{\mu}}{\mu}\frac{2\mu}{\hbar^{2}}E}{\frac{2\tilde{\mu}}{m_{2}}p_{3}^{2}-\frac{2\tilde{\mu}}{m_{2}}k_{1}p_{3}+\frac{\tilde{\mu}}{\mu}k_{F}^{2}-\frac{\tilde{\mu}}{\mu}\frac{2\mu}{\hbar^{2}}E}\right)\nonumber \\
 & \sim\frac{4}{\pi}\int_{k_{F}}^{\Lambda_{2}}dp_{3}\,\frac{p_{3}^{2}}{p_{3}^{2}-\frac{\tilde{\mu}}{\mu}\frac{2\mu}{\hbar^{2}}E}\nonumber \\
 & \approx\frac{4}{\pi}\left[\Lambda_{2}-k_{F}+\frac{k_{F}}{2}\ln\left(\frac{2k_{F}(\Lambda_{2}-k_{F})}{k_{F}^{2}-\frac{\tilde{\mu}}{\mu}\frac{2\mu}{\hbar^{2}}E}\right)\right],\label{omega_3}
\end{align}

\begin{align}
\Omega_{4}(E,g_{23}) & =\frac{4\pi\hbar^{2}}{2\tilde{\mu}g_{23}}+\frac{1}{2}\Omega_{3}(E).\label{omega_4}
\end{align}
In order that Eq. (\ref{rewritten sys of integral Eqs}) possesses
nontrivial solutions, it is required that

\begin{equation}
\Omega_{1}(E,g_{12};\tau=1)\Omega_{4}(E,g_{23})-\Omega_{2}(E)\Omega_{3}(E)=0,\label{condition for the solution}
\end{equation}
which gives rise to a relation between $g_{12}$ and $g_{23}$ through
$E$. Figure \ref{Fig_2} shows the result for $E\approx E_{\mathrm{thr}}.$

\subsection{{\normalsize{}Numerical solution of the system of two coupled integral
equations (\ref{Main Eq 1}) and (\ref{Main Eq 2})}}

As mentioned, we assume $F_{i}(\mathbf{k})=F_{i}(k)$, which implies
that we only consider the isotropic solutions of Eqs. (\ref{Main Eq 1})
and (\ref{Main Eq 2}). To solve Eqs. (\ref{Main Eq 1}) and (\ref{Main Eq 2})
numerically we therefore replace the three-dimensional integrals over
momenta by one-dimensional integrals over absolute values of each
momentum. We discretize each integral range such that the grid points
$\{x_{j}\}$, $j=1,2,\ldots,N$, are the set of zeros of the Legendre
polynomials $P_{N}(x)$. We approximate the integrals by a truncated
sum weighted by $w_{j}$:

\begin{equation}
w_{j}=\frac{2}{(1-x_{j}{}^{2})[P'_{N}(x_{j})]^{2}},\label{weights}
\end{equation}
where $P_{N}'(x)=dP_{N}(x)/dx$ \cite{Gaussian_quadrature_2,Gaussian_quadrature_3}.
This choice, which is the so-called Gauss-Legendre quadrature rule,
scales the range of integration from a given interval $(a,b)$ to
$(-1,1)$, and has order of accuracy exactly $2N-1$, which is the
highest accuracy among the other quadrature choices \cite{Gaussian_quadrature_1}.
We apply the Gauss-Legendre quadrature rule on each integral and construct
a matrix equation analog to each integral equation. For given values
of $E$ below $E_{\mathrm{thr}}$ we calculate the eigenvalues, which
then provide the corresponding values of the interaction parameter.
The functions $F_{1}$ and $F_{2}$ are also obtained as the eigenvectors
of the matrix equations. Notice that, due to the truncation on each
sum, the two-body continuum revealed in Figs. \ref{Fig_3}, \ref{Fig_4},
and \ref{Fig_5} has a finite range.

\subsection{{\normalsize{}Derivation and solution of Eq. (\ref{Eq for g23 zero}),
calculation of $g_{12}^{(\mathrm{c})}$, and derivation of Eq. (\ref{energy for high m1})}}

To derive Eq. (\ref{Eq for g23 zero}), recall that for vanishing
$g_{23}$ we have $F_{1}=0$, and therefore Eq. (\ref{Main Eq 2})
will have no effect anymore. As discussed in App. B, for a conventional
superconductor we make an approximation such that $k_{1}\sim0$ and
$k_{2}\sim k_{F}$. The system of three electrons is then described
by Eq. (\ref{Eq for g23 zero}) for $\tau=1$. Also, the system of
two electrons `1' and `2' (or `3') will be described by the
same equation when $\tau=1/2$. Equation (\ref{Eq for g23 zero})
for $\tau=1$ is then solved by

\begin{equation}
\Omega_{1}(E,g_{12};\tau=1)=0,\label{first approximation}
\end{equation}
where the function $\Omega_{1}(E,g_{12};\tau)$ was calculated in
App. B, see Eq. (\ref{omega_1}). Equation (\ref{first approximation})
provides now a relation between the interaction parameter $g_{12}$
and the energy $E$, see dashed red curves in Fig. \ref{Fig_4}. For
a system of two electrons `1' and `2' (or `3') we set $\tau=1/2$
and calculate the integral by the same argument. The lowest energy
two-body bound state is obtained by solving 

\begin{equation}
\Omega_{1}(E,g_{12};\tau=1/2)=0,\label{approximation for dimer}
\end{equation}
see solid red curves in Fig. \ref{Fig_4}. 

To calculate the onset of the trimer state analytically, we expand
Eq. (\ref{first approximation}) for $\Lambda_{2}-k_{F}\ll k_{F}$
and $\Lambda_{1}\ll k_{F}$ at the threshold energy $E_{\mathrm{thr}}=2E_{F}$,
and solve the leading order for the interaction parameter $g_{12}^{(\mathrm{c})}\equiv g_{12}(E=E_{\mathrm{thr}})$,
which results in Eq. (\ref{g_12 critical}).

To derive Eq. (\ref{energy for high m1}), we notice that the onset
of the trimer state for $m_{1}\gg m_{2}$ leads to the origin, $|g_{12}^{(\mathrm{c})}|\rightarrow0$,
cf. Eq. (\ref{g_12 critical}). In this case, to find the asymptotic
of the trimer we solve the integral appearing in Eq. (\ref{omega_1})
by changing variable $x\equiv p_{3}/k_{F}$. The integral bounds will
be $1$ and $\Lambda_{2}/k_{F}$. For a conventional superconductor
the upper bound $\Lambda_{2}/k_{F}$ is very close to the lower bound.
Therefore, we calculate the integral by making the leading order of
the integrand when $x\rightarrow1$. Solving the result for $E$ we
arrive at Eq. (\ref{energy for high m1}), see dashed green curve
in Fig. \ref{Fig_4} (c). 

\begin{thebibliography}{10}
\bibitem{units}In this Letter, momentum is measured in units of the
Fermi momentum and energy is measured in units of the threshold energy,
$E_{\mathrm{thr}}=2E_{F}=\frac{\hbar^{2}}{m_{2}}k_{F}^{2}$, where
$E_{F}$ is the Fermi energy.
\end{thebibliography}
\end{document}